\documentclass[english,aps,prl,reprint,superscriptaddress,longbibliography]{revtex4-2}

\usepackage[T1]{fontenc}
\usepackage[utf8]{inputenc}
\usepackage{amssymb}
\usepackage{graphicx}
\usepackage{subscript}
\usepackage{amsmath}
\usepackage[colorlinks=true,citecolor=blue,urlcolor=blue,linkcolor=blue]{hyperref}
\usepackage{nameref}
\usepackage[export]{adjustbox}
\usepackage{ragged2e}
\usepackage{comment}

\usepackage{subcaption}
\usepackage{caption}
\makeatletter
\usepackage{pslatex}
\newcommand{\figpart}[2]{Figure~\hyperref[#1]{\ref*{#1}#2}}
\usepackage{hyperref}
\usepackage{xcolor}
\hypersetup{
    colorlinks,
    linkcolor={blue!70!black},
    citecolor={blue!70!black},
    urlcolor={blue!70!black}
}

\usepackage{babel}

\newcommand{\X}{\tilde{X}^{2}\Sigma^{+}}
\newcommand{\A}{\tilde{A}^{2}\Pi_{1/2}}
\newcommand{\B}{\tilde{B}^{2}\Sigma^{+}}

\newcommand{\appropto}{\mathrel{\vcenter{
  \offinterlineskip\halign{\hfil$##$\cr
    \propto\cr\noalign{\kern2pt}\sim\cr\noalign{\kern-2pt}}}}}

\newcommand{\RFMOTinitialpower}{3.7} 
\newcommand{\RFMOTfinalpower}{0.5} 
\newcommand{\RFMOTinitialBgrad}{12} 
\newcommand{\RFMOTfinalBgrad}{16} 
\newcommand{\RFMOTinitialsize}{1.3} 
\newcommand{\RFMOTfinalsize}{650} 
\newcommand{\RFMOTramptime}{15} 
\newcommand{\RFMOTtemperature}{1} 

\newcommand{\LambdaTime}{2} 

\newcommand{\LambdaInt}{3.7} 
\newcommand{\LambdaDetuning}{13.2} 
\newcommand{\LambdaTwoPhoton}{-0.4} 

\newcommand{\LambdaTemp}{34} 
\newcommand{\LambdaTempUnc}{3} 

\newcommand{\SFtemp}{16.9} 
\newcommand{\SFtempunc}{1.3} 
\newcommand{\SFInt}{9.5} 
\newcommand{\SFdet}{96} 
\newcommand{\SFtime}{5} 

\newcommand{\ConveyorRamptime}{10} 
\newcommand{\ConveyorBgrad}{88} 

\newcommand{\ConveyorIntBefore}{14.0} 
\newcommand{\ConveyorIntAfter}{8.9} 

\newcommand{\ConveyorSize}{83} 
\newcommand{\ConveyorSizeUnc}{1}

\newcommand{\ConveyorTemp}{91} 
\newcommand{\ConveyorTempUnc}{4} 

\newcommand{\ConveyorDetuning}{7.7} 
\newcommand{\ConveyorDeltaA}{-0.9} 
\newcommand{\ConveyorDeltaB}{0.1} 
\newcommand{\ConveyorCompression}{245} 

\newcommand{\ODTSize}{50} 

\newcommand{\ODTPower}{10.8} 
\newcommand{\ODTTrapDepth}{750}

\newcommand{\ODTnumber}{1400}
\newcommand{\ODTnumberUnc}{300}
\newcommand{\ODTHoldTime}{30}
\newcommand{\ODTImageTime}{50}

\newcommand{\ODTLifetimeMain}{1.5} 
\newcommand{\ODTLifetimeMainunc}{0.1} 
\newcommand{\ODTLifetimeEDM}{320} 
\newcommand{\ODTLifetimeEDMunc}{30} 
\newcommand{\ODTLifetimeDMStretch}{135} 
\newcommand{\ODTLifetimeDMStretchunc}{17} 
\newcommand{\ODTLifetimeDMBend}{190} 
\newcommand{\ODTLifetimeDMBendunc}{30} 

\newcommand{\TisapphPowerUDM}{100}

\newcommand{\DyePowerUDM}{100}

\makeatother

\begin{document}

\title{Optical Trapping of SrOH Molecules for Dark Matter and T-violation Searches}

\author{Hiromitsu Sawaoka}
\altaffiliation{Current address: Department of Physics, University of California, Berkeley, CA 94720, USA}
\affiliation{Harvard-MIT Center for Ultracold Atoms, Cambridge, Massachusetts 02138, USA}
\affiliation{Department of Physics, Harvard University, Cambridge, Massachusetts 02138, USA}

\author{Abdullah Nasir}
\affiliation{Harvard-MIT Center for Ultracold Atoms, Cambridge, Massachusetts 02138, USA}
\affiliation{Department of Physics, Harvard University, Cambridge, Massachusetts 02138, USA}

\author{Annika Lunstad}
\affiliation{Harvard-MIT Center for Ultracold Atoms, Cambridge, Massachusetts 02138, USA}
\affiliation{Department of Physics, Harvard University, Cambridge, Massachusetts 02138, USA}

\author{Mingda Li}
\affiliation{Harvard-MIT Center for Ultracold Atoms, Cambridge, Massachusetts 02138, USA}
\affiliation{Department of Physics, Harvard University, Cambridge, Massachusetts 02138, USA}

\author{Jack Mango}
\affiliation{Harvard-MIT Center for Ultracold Atoms, Cambridge, Massachusetts 02138, USA}
\affiliation{Department of Physics, Harvard University, Cambridge, Massachusetts 02138, USA}

\author{Zack D. Lasner}
\altaffiliation{Current address: IonQ, Inc., College Park, MD 20740, USA}
\affiliation{Harvard-MIT Center for Ultracold Atoms, Cambridge, Massachusetts 02138, USA}
\affiliation{Department of Physics, Harvard University, Cambridge, Massachusetts 02138, USA}

\author{John M. Doyle}
\affiliation{Harvard-MIT Center for Ultracold Atoms, Cambridge, Massachusetts 02138, USA}
\affiliation{Department of Physics, Harvard University, Cambridge, Massachusetts 02138, USA}

\date{\today}

\begin{abstract}

We report an optical dipole trap of strontium monohydroxide (SrOH) with 1400(300) trapped molecules.
Through optical pumping, we access 
vibrational states that are proposed for improved probes of the electron's electric dipole moment (eEDM)  and ultralight dark matter (UDM)~\cite{KozyryevPolyEDM, kozyryev2021enhanced}. For each of these states, the lifetime of trapped molecules is measured, and found to be consistent with spontaneous radiative decay and black-body excitation limits, making this platform viable for these eEDM and UDM searches.

\end{abstract}

\maketitle

\emph{Introduction}---Molecules, due to their rich internal structure, are useful for many applications in quantum science, including quantum information~\cite{demille2002quantum, yelin2006schemes, ni2018dipolar, sawant2020ultracold, wei2011entanglement, yu2019scalable, albert2020robust}, quantum chemistry and collisions~\cite{heazlewood2021towards}, and precision searches of physics beyond the Standard Model (BSM)~\cite{KozyryevPolyEDM, kozyryev2021enhanced, norrgard2019nuclear, kobayashi2019measurement}. To take full advantage of these applications, trapped ultracold molecules and realization of long internal quantum state coherence times are of key importance~\cite{BaoDipolarSpinExchange, HollandOnDemandEntanglement, VilasCaOHTweezerArray, AndereggPathfinder}. Towards these ends, recent laser cooling and optical pumping experiments 
have yielded several species of diatomic and polyatomic molecules prepared with single quantum state control in the ultracold regime~\cite{AndereggCaF, WilliamsCaF, BarrySrF, CollopyYO, ZengBaF, Padilla-Castillo_AlF, VilasCaOH, LasnerMOT, Prehn2016}. 

The rotational degree of freedom that is present in all molecules is a resource for quantum science. Generically, polyatomic molecules have an additional useful property, the projection of angular momentum along the internuclear axis. This degree of freedom is a powerful tool for precision measurement~\cite{KozyryevPolyEDM, kozyryev2021enhanced, norrgard2019nuclear, hutzler2020polyatomic} and has 
potential for quantum simulation~\cite{Wall2013, Wall2015, yu2019scalable}. Linear combinations of these projection states give rise to long-lived parity doublets (PDs). PD states can be fully polarized using only very modest external electric fields, resulting in optically addressable oppositely oriented molecules. PDs also have specific structural features that aid in suppression of systematic errors, as used in the current leading electron electric dipole moment (eEDM) searches~\cite{AndreevACME, JilaEDM}. 

A certain class of heavy polyatomic molecules, which includes strontium monohydroxide (SrOH), presents important opportunities because they are amenable to both laser cooling and optical trapping, while also having long-lived PD states. The $\X (010)$ \footnote{vibrational states of linear triatomic molecules such as SrOH are denoted as $(v_1 {v_2}^{\ell} v_3)$, where $v_1$, $v_2$, and $v_3$ correspond to quanta in the symmetric stretch, bend, and antisymmetric stretch modes, respectively. The quantum number $\ell$ represents the vibrational angular momentum associated with the bending mode and is typically omitted when $v_2 = 1$, where $\ell = 1$ is the only allowed value.} vibrational bending mode in SrOH has such PD states. This makes it a promising candidate for future eEDM measurements that can enjoy the very long coherence times available with optically trapped molecules~\cite{KozyryevPolyEDM, AndereggPathfinder}. SrOH can also be used for sensing variations in fundamental constants, another frontier of BSM physics searches. In well-motivated models of dark matter, oscillations of the ultralight dark matter (UDM) field can couple to fundamental constants, such as the proton-to-electron mass ratio $\mu \equiv m_p/m_e$ ~\cite{arvanitaki2015searching, stadnik2015can, graham2013new, brdar2018fuzzy, banerjee2019coherent, stadnik2016improved, brzeminski2021time, cosme2018scale}. The gap between pairs of states within a group of closely spaced heterogeneous vibrational levels in SrOH -- between a rotational state in the $\X (200)$ manifold and a rotational state in the $\X (03^10)$ manifold -- varies sensitively with changes of $\mu$ due to the differing anharmonicities of these vibrational modes. The transitions arising from these $\X(200)-\X(03^10)$ pairs 
have the technical advantage of being in the microwave regime. Due to the many rotational states in each vibrational manifold, multiple combinations of UDM sensitive states can be easily studied for systematic error rejection~\cite{kozyryev2021enhanced}.

In this Letter, we report optical trapping of SrOH. We achieve this through sub-Doppler cooling~\cite{Truppe2017CaFSubDoppler, CheukCaFLambdaGMC, DingYOGMC, HallasOpticalTrappingCaOH}, spatial compression with a conveyor-belt (CB) MOT~\cite{LiConveyorBeltMOT, HallasBlueMOT, YuConveyorBeltMOTCaF, ZengBaFBlueDetunedMOT}, and single-frequency cooling to load an optical dipole trap (ODT).~\cite{Caldwell2019_DeepCoolingMolecules, HallasOpticalTrappingCaOH}. 
We conduct molecular lifetime measurements of selected vibrational states in the ODT. The selected states are sensitive to the eEDM and to variations of $\mu$.
These results demonstrate trapped ultracold SrOH for improved searches for new time-reversal violating physics (T-violation) in the $>10$~TeV mass range~\cite{KozyryevPolyEDM} and UDM searches in the $10^{-22}$ to $10^{-14}$ eV mass range~\cite{kozyryev2021enhanced}.

\begin{figure*}[t]
  \centering
  \includegraphics[width=\textwidth]{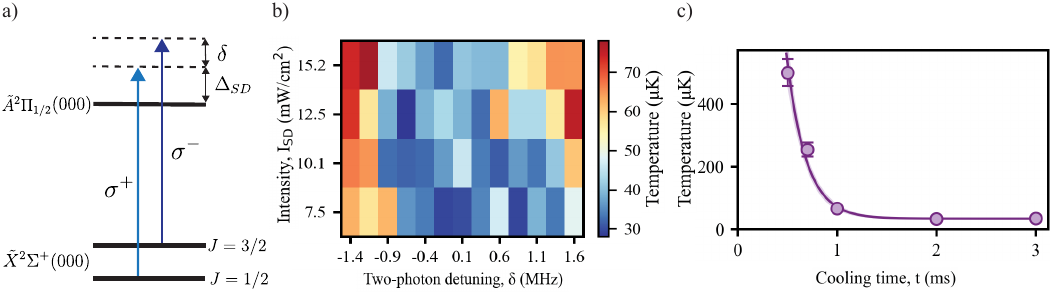}

  \captionsetup{justification=justified, singlelinecheck=false}
  \caption{\justifying Sub-Doppler cooling of SrOH. (a) Level diagram showing transitions involved in $\Lambda$-cooling. For SF cooling, only the frequency addressing the $J = 1/2$ ground state (with $\sigma^+$ polarization) is present.
           (b) temperature (T) of the SrOH molecules depending on two-photon detuning $\delta$ and total cooling light intensity $I_{SD}$.
           (c) T vs cooling time (the time over which cooling light is present) under optimal $\Lambda$-cooling parameters, where an exponential decay function ($e^{-t/\tau_{\Lambda}}$) is fit, yielding a characteristic time of $\tau_\Lambda = 0.19(1)$~ms. For later times, error bars are smaller than the points.}
  \label{fig:Lambda}
  \label{fig:Lambda:a}
  \label{fig:Lambda:b}
  \label{fig:Lambda:c}
\end{figure*}

\emph{Trapping and Sub-Doppler Cooling}---Magneto-optical trapping (MOT) is a workhorse technique used to reach the ultracold regime. We previously demonstrated a radio-frequency MOT (RF MOT) of SrOH, using (typical) red detuning  
from the main transition, $\X(000) \rightarrow \A(000)$.
Details of the laser cooling scheme, slowing technique, and RF MOT 
can be found in \cite{LasnerMOT}. 
In our current work, following initial capture of molecules in the RF MOT, the intensity of the RF MOT light beams are lowered from \RFMOTinitialpower  $\text{ mW/cm}^2$ to \RFMOTfinalpower  $\text{ mW/cm}^2$, and the magnetic field gradient is increased from \RFMOTinitialBgrad~G/cm to \RFMOTfinalBgrad~G/cm, over $\RFMOTramptime$~ms. This spatially compresses the RF MOT cloud of molecules from an initial root-mean-squared diameter of $\sigma =\RFMOTinitialsize$~mm down to $\sigma = \RFMOTfinalsize$~$\mu$m, with a molecular temperature of $\sim\RFMOTtemperature$~mK. 

Efficient loading of an ODT requires molecules that are much colder than the ODT trap depth, in our case $\sim \ODTTrapDepth$~$\mu$K. We achieve this with sub-Doppler cooling using blue-detuned optical molasses methods \cite{Truppe2017CaFSubDoppler, VilasCaOH}. 
To preserve rotational closure in molecules, 
"type-II" laser cooling schemes are used, where the angular momentum in the ground manifold exceeds that of the excited manifold. As such, dark states are present in the ground manifold, leading to polarization gradient forces that provide sub-Doppler cooling (heating) for blue (red) detuning relative to the main transition \cite{BoironSD, DevlinSD, SieversSD}. 

Two sub-Doppler configurations are used in our work, "$\Lambda$-cooling" and "SF Cooling", as described below. Experimentally, sub-Doppler cooling is achieved in the following way. First, the RF MOT magnetic field and light polarization switching are both turned off. Then, to form a blue detuned molasses, the detuning of the trapping laser is rapidly changed from $\sim 1\text{ } \Gamma$  (7~MHz) red of the main transition to a blue detuning $\Delta_{SD}$. The realized molasses consists of two frequency components addressing the spin-rotation components $J = 1/2$ and $J = 3/2$ of the ground state $\X(000;N=1)$, with $\sigma^+$ and $\sigma^-$ polarizations, respectively, up to the excited state $\A(000;J=1/2)$. Appropriately tuning the two-photon detuning $\delta$ (see \figpart{fig:Lambda}{a}) creates zero-velocity dark states and provides enhanced cooling through velocity-selective coherent population trapping~\cite{AspectVSCPT}. This configuration is known as 
$\Lambda$-enhanced gray molasses cooling ($\Lambda$-cooling) \cite{CheukCaFLambdaGMC, LanginLambda}. We use the high capture velocity of $\Lambda$-cooling to cool the molecules directly from MOT temperatures of $\sim1$~mK.
$\Lambda$-cooling light is applied for $\LambdaTime$~ms, longer than the characteristic cooling time $\tau_{\Lambda} = 0.19(1)$~ms, see \figpart{fig:Lambda}{c}. The resultant temperature of the molecular cloud is measured using time-of-flight expansion, as described in the Supplemental Material.  A minimum temperature of $\LambdaTemp(\LambdaTempUnc)$~$\mu$K is achieved with $\Delta_{SD} = \LambdaDetuning$~MHz, $\delta = \LambdaTwoPhoton$~MHz, and total light beam intensity $I_{SD} = \LambdaInt$~$\text{mW/cm}^2$.

Single frequency (SF) cooling \cite{Caldwell2019_DeepCoolingMolecules} is used to directly load the ODT, as described later, after the section on the Conveyor-belt MOT. SF cooling achieves lower temperatures than $\Lambda$ cooling and is less affected by the ODT trap light shifts. However, because its capture velocity is too low to directly cool from the RF MOT, we must first apply $\Lambda$-cooling. Our SF cooling employs only the frequency component that addresses the $J = 1/2$ spin-rotation state (i.e., the $J = 3/2$ light is simply turned off). We set $\Delta_{SD} = \SFdet$~MHz and $I_{SD} = \SFInt$~$\text{mW/cm}^2$, for both ODT loading and for free space cooling studies. With cooling light applied for $\SFtime$~ms, a minimum temperature of $\SFtemp(\SFtempunc)$~$\mu$K is achieved in free space. 

\begin{figure*}[t]
  \centering
  \includegraphics[width=\textwidth]{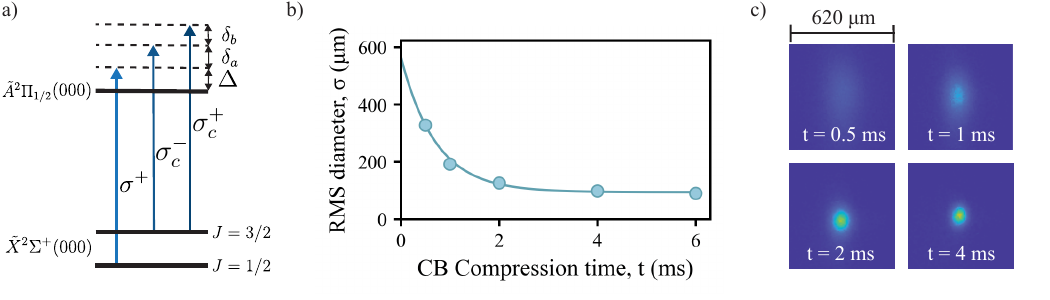}

  \captionsetup{justification=justified, singlelinecheck=false}
  \caption{\justifying Conveyor belt MOT (CB MOT) of SrOH. (a) Level diagram showing transitions involved in the CB MOT,
           where the additional frequency component $\sigma_c^+$ is introduced after $\Lambda$-cooling, forming the conveyor belt condition.
           (b) CB MOT spatial RMS diameter $\sigma$ as a function of compression time,
           where the data are fit to a decaying exponential function ($e^{-t/\tau_{CB}}$) resulting in $\tau_{CB} =0.7(1)$ ms. The $2\sigma$ error bars are smaller than the heights of
           the data points, and are not visible.
           c) EMCCD fluorescence images of molecules compressed in the CB MOT at different compression times $t$. At $t = 4$~ms, the density of the cloud is increased by a factor of $\ConveyorCompression$.}
  \label{fig:Conveyor}
  \label{fig:Conveyor:a}
  \label{fig:Conveyor:b}
  \label{fig:Converoy:c}
\end{figure*}

\emph{Conveyor-belt MOT}---Once the molecules are cooled to sub-Doppler temperatures using the $\Lambda$-cooling described above, a key remaining step towards efficient ODT loading (with SF cooling) is further compression of the molecular cloud. Given that the RF MOT has a final size of $\sigma = \RFMOTfinalsize$~$\mu$m compared to the ODT waist diameter of $\sigma_{ODT} = \ODTSize$~$\mu$m, direct loading from the RF MOT is inefficient due to the large spatial mode mismatch. The advent of the conveyor-belt MOT (CB MOT) \cite{LiConveyorBeltMOT, YuConveyorBeltMOTCaF, HallasBlueMOT} allows for increased compression, with a resulting $\sim$10 fold decrease in cloud diameter. As described in Ref \cite{LiConveyorBeltMOT}, this 
arises from a slow moving lattice with molasses that shuttles the trapped molecules towards the magnetic field center where they eventually settle.

As shown in \figpart{fig:Conveyor}{a}, the CB MOT uses a ``(1+2)'' frequency configuration, where one frequency component addresses the $J = 1/2$ spin-rotation level with $\sigma^+$ polarization and two other components address the $J = 3/2$ level, with $\sigma^-$ and $\sigma^+$ polarizations. The CB MOT beams share a common blue-detuning $\Delta_{CB}$, and we refer to their relative two-photon detunings as $\delta_a$ and $\delta_b$. Experimentally, the CB MOT immediately follows $\Lambda$-cooling -- a DC MOT magnetic field is turned on and the additional light frequency ($\sigma_c^+$) is added. To achieve optimal compression, the DC magnetic field gradient, $B'_{CB}$, is subsequently ramped up and the total CB MOT light intensity, $I_{CB}$, is ramped down, over a time $\tau_{CB}$. 

We explore the dependence of the CB MOT size $\sigma$ on the various detunings, beam intensities, ramp time, and magnitude of the magnetic field gradient. We achieve a temperature of $\ConveyorTemp (\ConveyorTempUnc)$~$\mu$K by employing $\Delta_{CB} = \ConveyorDetuning$~MHz, $\delta_a = \ConveyorDeltaA$~MHz, $\delta_b = \ConveyorDeltaB$~MHz, and with the light beam intensities ramped from $\ConveyorIntBefore$$ ~\text{mW/cm}^2$ to $\ConveyorIntAfter$$ ~\text{mW/cm}^2$ and a final magnetic field gradient $B'_{CB} = \ConveyorBgrad$~G/cm. The CB MOT is compressed over a time $\tau_{CB} = \ConveyorRamptime$~ms, with $\sim50\%$ of the molecules remaining in the trap. Starting from the RF MOT cloud diameter $\sigma = \RFMOTfinalsize$~$\mu$m, CB MOT compression results in a molecular cloud diameter of $\sigma =\ConveyorSize (\ConveyorSizeUnc)~\mu$m. This cloud size is well matched to the diameter of our ODT light beam, $\sigma_{ODT} = \ODTSize~\mu$m. 

\emph{Optical Dipole Trapping}---After cooling and compression in the CB MOT, SrOH molecules are loaded into an ODT \cite{HallasOpticalTrappingCaOH} using SF cooling. The ODT is formed by  a focused 1064~nm beam 
with $\ODTPower$~W of power,  producing a trap depth of $\sim \ODTTrapDepth$~$\mu$K. To load molecules into the ODT we turn on SF cooling for 60~ms with the ODT light on.  SF cooling is advantageous for this stage as the light shift due to the infrared ODT beam does not appreciably change the optimal cooling detuning. Once the molecules are loaded, the trap is held for a total of $\ODTHoldTime$~ms with all cooling lasers switched off. The molecules are then imaged with $\Lambda$-cooling light for $\ODTImageTime$~ms. Employing an EMCCD camera, we detect the 611~nm wavelength photons emanating from the decay of molecules driven through the $\B(000)$ state.

We observe a total number of $N = \ODTnumber(\ODTnumberUnc$) SrOH molecules trapped in the ODT. To measure this quantity, the molecules are recaptured in the RF MOT after being held in the optical trap, and the number of molecules is determined using the procedure detailed in \cite{LasnerMOT}. 

\begin{figure}[t]
  \noindent
  \begin{minipage}[t]{\columnwidth} 
    \centering
    \includegraphics[width=\linewidth]{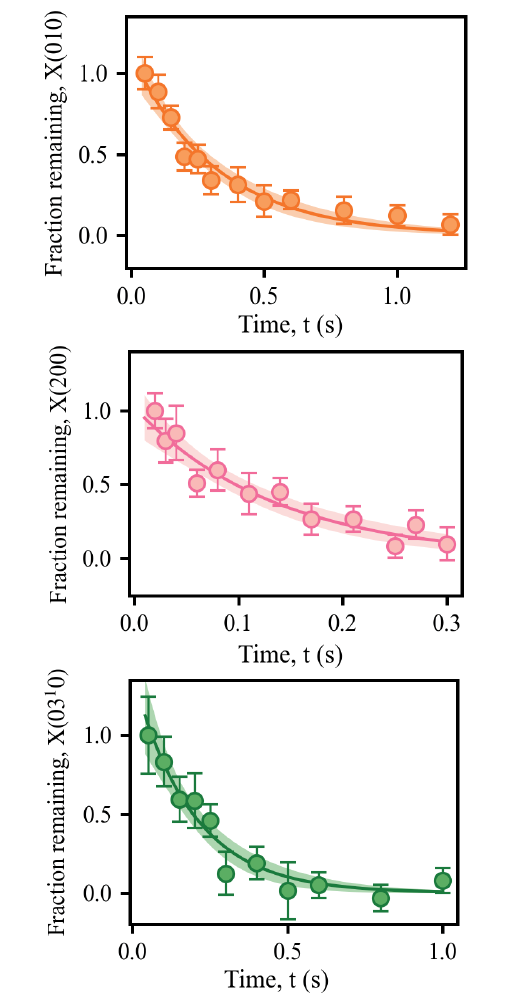}\par\vspace{3mm}

    \captionsetup{width=\linewidth, justification=justified, singlelinecheck=false}
    \captionof{figure}{\justifying Science state lifetimes in the ODT. Lifetimes are extracted by measuring the molecular fluorescence over time $t$, and fitting to a decaying exponential function. Error bars are $1\sigma$, and the shaded area around curves correspond to the $1\sigma$ error on the fit. The measured lifetimes are $\ODTLifetimeEDM(\ODTLifetimeEDMunc)\text{, } \ODTLifetimeDMStretch(\ODTLifetimeDMStretchunc)\text{, and }  \ODTLifetimeDMBend(\ODTLifetimeDMBendunc)$~ms, for the $\X(010),~\X(200),\text{ and }\X(03^10)$ states, respectively.}
    \label{fig:ODT_lifetimes}
  \end{minipage}
\end{figure}

\emph{Science State Lifetimes in the ODT}---We measure the lifetimes of molecules in the ODT for the vibrationally excited ``science states'' $\X(010)$, $\X(200)$ and $\X(03^10)$.
The molecules are first prepared in the ODT in the state, after which they are held for varying times (up to $T =$ 2 s). The molecule number is measured as a function of $T$ and we fit a decaying exponential to extract the time constant corresponding to the lifetime of the molecules in the trap, see Figure \ref{fig:ODT_lifetimes}. 

We also measure the lifetime of molecules in the ODT without further state preparation after SF cooling, which leaves the population dominantly in the ground state $\X(000)$. The lifetime is $\tau = \ODTLifetimeMain(\ODTLifetimeMainunc)$~s,
which is dominated by 
black-body radiation (BBR) excitation driven loss
of $\X(000)$ molecules (see Supplemental Material).

The $\X(010)$ state has been proposed for an eEDM measurement \cite{KozyryevPolyEDM, hutzler2020polyatomic, AndereggPathfinder} due to its closely spaced parity doublet structure and good sensitivity~\cite{Gaul2020}. Since this state is addressed in our optical cycle, we
optically pump into the state by simply turning off the corresponding repump laser that addresses the $\X(010)\rightarrow \B(000)$ transition. 
We measure a lifetime of $\X(010)$ molecules in the ODT of $\tau_{(010)} = \ODTLifetimeEDM(\ODTLifetimeEDMunc)$~ms. 

The $\X(200)$ and $\X(03^10)$ states have been proposed for UDM searches \cite{kozyryev2021enhanced}, where the energy of the $\X(03^10)$ state has recently been measured to high precision \cite{Lunstad2025}. The method to prepare the molecules in $\X(200)$ is similar to that for $\X(010)$, given that it also resides in our optical cycle. 
We measure a lifetime $\tau_{200} = \ODTLifetimeDMStretch(\ODTLifetimeDMStretchunc)$~ms. For $\X(03^10)$, we must incorporate a separate optical pumping step because the state is not present in the optical cycle. This is done by first populating the $\X(010)$ state as described above. Next, the $\X(010) \rightarrow \tilde{A}(030)\kappa^2\Pi_{1/2}$ transition is driven 
using $\sim\DyePowerUDM$~mW of laser power. 
The $\tilde{A}(030)\kappa^2\Pi_{1/2}$ state predominantly decays to the $\X(03^10)$ ground state. In order to detect the molecules, they must be transferred back into the optical cycle. This is done by driving the $\X(03^10) \rightarrow \tilde{A}(010)\kappa^2\Sigma_{1/2}$ transition 
using a 
\linebreak
{$\sim\TisapphPowerUDM$~mW} light beam.
We report the lifetime $\tau_{(03^10)} = \ODTLifetimeDMBend(\ODTLifetimeDMBendunc)$ ms. 

All of these science state molecular lifetimes in the ODT  are consistent, within reasonable uncertainties, with the theoretically determined decay rates (see Supplemental Material), including both spontaneous and BBR driven loss.


\emph{Conclusion}---In summary, we demonstrate an optical dipole trap (ODT) of SrOH molecules and prepare the trapped molecules 
in the  $\X(010)$, $\X(200)$ and $\X(03^10)$  science states, which are proposed for searches of BSM physics.  
The lifetimes of molecules in the ODT are dominantly limited only by spontaneous radiative decay and black-body radiation excitation loss. 
The long interrogation times available in this system, on the order of hundreds of milliseconds, makes this a promising platform for EDM, UDM, and other future precision measurements. 
With the demonstrated number of $10^3$ trapped SrOH molecules in the ODT, improvement in the search for UDM over the current best limits is available, as proposed in Ref \cite{kozyryev2021enhanced}.

In the future, through application of transverse cooling ~\cite{Alauze_2021_YbF_transverse, Langin_2023_improved_loading} of the CBGB beam~\cite{HutzlerCBGB} that feeds the RF MOT, plus smaller technical improvements, we estimate that trapping $10^5$ molecules in the ODT is possible.
With this projected number of molecules and the near unity state preparation efficiency, an eEDM measurement with competitive sensitivity to the next projected ACME result should be possible using the $\X(010)$ bending mode of SrOH. Methods to conduct such eEDM measurements in polyatomic molecules have been demonstrated in CaOH \cite{AndereggPathfinder}.  In searching for the eEDM, it is important to have experiments with very different possible systematic errors, as is the case with the two best published limits, the ACME and JILA experiments ~\cite{AndreevACME,JilaEDM}. The work here demonstrates the key experimental building block for a viable third experimental approach to eEDM searches and provides a milestone towards trapping of more exotic species, such as RaOH~\cite{zhang2023relativistic, Conn2025}, which can provide even further improvements in the search for new T-violating physics, into the 1000 TeV range.

\emph{Acknowledgments}---We are grateful to Nick Hutzler for his extensive help during this work and for important discussions and feedback on the manuscript. We thank Lan Cheng and Chaoqun Zhang for calculations necessary to estimate the theoretical lifetimes of the science states. We additionally thank Christian Hallas and Grace Li for their assistance and expertise with the conveyor-belt blue-detuned MOT. We also thank Paige Robichaud, Nathaniel Vilas, and Avikar Periwal for helpful discussions, and additionally thank Nathaniel for feedback on the manuscript.

This work was done at the Center for Ultracold Atoms (an NSF Physics Frontier Center) and supported by Q-SEnSE: Quantum Systems through Entangled
Science and Engineering (NSF QLCI Award OMA-2016244), the Alfred P. Sloan Foundation (G-2023-21036), the Gordon and Betty Moore
Foundation (7947), AOARD: Asian Office of Aerospace Research and Development  (FA2386-24-1-4070), and AFOSR: Air Force Office of Scientific Research (DURIP FA9550-24-1-0060).



\bibliographystyle{apsrev4-2}
\bibliography{sroh}

\begin{center}
\textbf{Supplemental Material}
\end{center}

\appendix

\section{Sub-Doppler temperature measurements}
\label{SM:Tempmeasurements}

In order to measure the temperature of the sub-Doppler cooled cloud of SrOH molecules, we use the standard ballistic expansion method. After sub-Doppler cooling, the cloud of molecules is allowed to ballistically expand for a variable amount of time, $t$, with all repumping and cooling lasers switched off. Next, the $X^2\Sigma^+(000) \to A^2\Pi_{1/2} (000)$ mainline transition is resonantly driven, as well as all other repumping transitions for 2 ms. During this time, 611 nm photons from the vibrationally diagonal $B^2\Sigma^+\to X^2\Sigma^+$transitions are detected on an EMCCD camera \cite{LasnerMOT}. The root mean square (RMS) widths of the cloud in the radial and axial directions, $\sigma_{\rho}$ and $\sigma_z$ respectively, are extracted by fitting an elliptical Gaussian to an averaged 2D image for each expansion time. The radial and axial temperatures are determined by fits to the linearized ballistic expansion equation, $$\sigma_i^2 = \sigma_{i, 0}^2 + k_B T_i t^2 / m$$ where $\sigma_{i, 0}$ is the initial width, $k_B$ is the Boltzmann constant, $m$ is the mass of an SrOH molecule, and $T_i$ is the temperature which we aim to extract. Finally, an  ``overall temperature'' $T_o$ is reported as the weighted geometric mean of radial and axial temperatures $T_o=T_{\rho}^{2/3}T_{z}^{1/3}$.  

\begin{figure}[h]
  \centering
  \includegraphics[width=0.9\columnwidth]{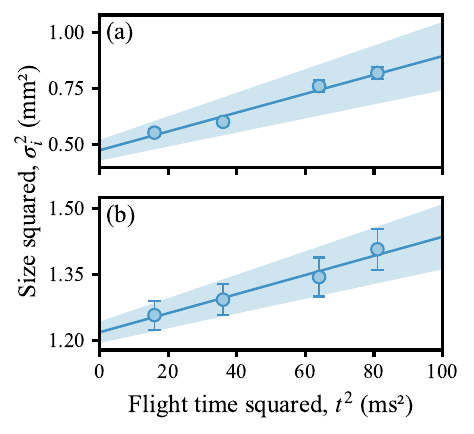}

  \captionsetup{justification=justified, singlelinecheck=false}
  \caption{\justifying Time of flight temperature measurement of the sub-Doppler cooled cloud of SrOH molecules. Parameters for cooling are $\Delta=13.2$ MHz, $\delta=-0.4$ MHz, and $I_{SD} = 11.8~\text{mW/cm}^2$. Expansion times are 4, 6, 8, and 9 ms. The fitted temperatures in the axial (a) and radial (b) direction are $T_z=52(7)\mu$K and $T_\rho=27(3)\mu$K, respectively, giving a geometrically averaged (``overall'') temperature of $T_o=33(3)\mu$K. The shaded regions around the linear fits corresponds to $\pm 1\sigma$ error bounds.}
  \label{fig:tof-measurement-sup}
\end{figure}

During imaging, the molecule cloud continues to expand and is heated due to photon scattering. To avoid these unwanted systematic errors from biasing temperature measurements, the imaging time must be sufficiently short compared to the expansion time. We measure the effect of imaging time on the apparent width of the molecule cloud, comparing imaging times ranging between 0.5 ms and 4 ms as shown in Figure~\ref{fig:tof-measurement-sup}. We find there is no significant increase in molecule cloud size when imaging for as long as 2 ms, compared to shorter times. 



\section{ODT setup}

The ODT light is generated by a 50~W PreciLaser 1064~nm fiber amplifier seeded by a RIO ORION 1064~nm laser, selected for its low relative intensity noise (RIN). A Faraday isolator is placed directly after the laser head to suppress back-reflections. The beam then passes through an acousto-optic modulator (AOM, Gooch \& Housego 3080-199), which is used for active power stabilization via an analog PID controller (SRS SIM960).

The AOM output is coupled into a photonic crystal fiber (PCF, LMA-PM-15) to improve beam quality and pointing stability. The fiber output is collimated with an aspheric lens (C280TME-1064), expanded by a 2$\times$ telescope (GBE02-C), and focused into the MOT chamber with an achromatic lens. This beam waist of the ODT beam is $w_0= 25~\mu m$.

In normal operation, approximately 10.8~W reaches the chamber, yielding a trap depth of $\sim 750~\mu\mathrm{K}$. With our sub-Doppler cooling temperatures, this corresponds to a trap parameter of $\eta \approx 20$ for SrOH, where $\eta$ is the ratio of the trap depth to the temperature of the molecules.

\section{Imaging the ODT}

We use two complementary methods to image molecules in the ODT: $\Lambda$-cooling imaging and RF MOT recapture.


To directly image trapped molecules we use a similar scheme to that described in appendix A. The blue detuned $\Lambda$-cooling configuration is used rather than the resonant mainline transition. Molecules are imaged with the trap light on for 50~ms. The resulting fluorescence provides position and spatial-distribution information while maintaining low temperature. However, due to lower scattering rate, this method is not suitable for precise molecule-number measurements.


For precise molecule-number measurements, e.g. in the ODT, we recapture into the RF MOT, which offers a much higher scattering rate than $\Lambda$-cooling. After initiating ODT loading, we wait 100~ms before switching on the RF MOT laser beams, to allow untrapped molecules to leave the capture region. We then turn off the ODT light and image the recaptured cloud in the RF MOT for 100~ms, sufficient to reach the full molecule photon budget. Using the known photon budget, the branching ratio into the detected channel, the geometric collection efficiency of our imaging system, and the camera’s signal-to-count calibration, we extract the trapped molecule number (see Ref.~\cite{LasnerMOT}).

\section{
Lifetimes of the Science States}\label{SM:Lifetime estimates}

{\bf{Experiment}}---The ODT lifetimes of the science states are measured by first pumping the trapped molecules into the science state, i.e. the ``target state'' (with at least 50\% pumping efficiency), and then pushing out from the ODT all of the molecules that were not prepared in the target state. This is done by turning on the light used for slowing the CBGB beam (which also passes through the MOT region), for $\sim$6~ms, holding the prepared molecules in the ODT for variable times, pushing out all molecules that might have transitioned out of the target state again, and then imaging the remaining molecules  with $\Lambda$-cooling. The push out is necessary to ensure that we do not falsely measure molecules that have decayed out of the target state to another state in the optical cycle. 
We fit the imaged ODT to an exponential decay function. 

{\bf{Theory}}---The decay of molecules from the ODT naturally includes loss due to both the spontaneous radiative decay and the scattering of environmental black-body photons, called black-body radiation (BBR) loss. The latter depends on the ambient temperature of the environment. The spontaneous decay rate between states $i$ and $j$  given by 
\begin{equation*}
    \Gamma_{\text{sp},ij }= \frac{\omega_{ij}^3 }{3\pi\epsilon_0\hbar c^3}S_{ij},
\end{equation*}
where $\omega_{ij}$ is the energy difference between $i$ and $j$ and $S_{ij}$ is the transition strength between those states, which can be derived using the transition dipole moment \cite{Vilas2023_BlackbodyLifetimes}. Knowledge of the transition dipole moment requires calculations of the potential energy surfaces for the vibrational mode and as such, is computationally intensive. Once the dipole moment is known, the spontaneous lifetime $\Gamma_\text{sp}$ can be calculated and used to determine the lifetime due to BBR induced decays. The BBR decay rate is calculated by summing over possible decays from state $i\rightarrow j$ and is given by 
\begin{equation*}
    \Gamma_{\text{BBR},ij} =  \sum_{i\neq j}  \Gamma_{\text{sp}, i\rightarrow j} \frac{1}{e^{\hbar \omega_{ij}/(k_BT)}-1},
\end{equation*}
 where $k_B$ is the Boltzmann constant, $T$ is the temperature of the environment, and $\omega_{ij}$ is the energy difference between states $i$ and $j$. 

The term $S_{ij}$ has contributions from both rotational and vibrational transitions and is given as $S_{ij}=S_{ij}^\text{vib}S_{ij}^\text{rot}$. In this work, we ignore the rotational contributions as our calculations have shown them to be constant among all transitions we consider. Then, $S_{ij}^\text{vib}$ is given as
\begin{align}
    S_{ij}^{vib.} &= \lvert \langle v_{1i}, v_{2i}, \ell_i,0 |\vec\mu |  v_{1j}, v_{2j}, \ell_j,0\rangle \rvert ^2 \notag \\
    &\approx \left| \frac{d\vec\mu}{dQ_1}  \langle v_{1i}|Q_1|v_{1j} \rangle  +  \frac{d\vec\mu}{dQ_2} \langle v_{2i},\ell_i|Q_2|v_{2j},\ell_j \rangle \right|^2,
    \label{eq:Sij}
\end{align}
where $Q_{1,2}$ are the normal vibrational coordinates for $v_{1,2}$, respectively. In this treatment, we use the 2D harmonic oscillator approximation. Specifically, the matrix elements are given by
\begin{align*}
    |\langle v_1+1|Q_1|v_1\rangle|^2
        &= \tfrac{1}{2}(v_1+1) \\
    |\langle v_2+1,\ell\pm1|Q_2|v_2,\ell\rangle|^2
        &= \tfrac{1}{4}\bigl(1+\delta_{\ell,0}+\delta_{\ell\pm1,0}
           -\delta_{\ell,0}\delta_{\ell\pm1,0}\bigr) \\
        &\qquad\times\left(\tfrac{v_2\pm\ell}{2}+1\right).
\end{align*}
With this approximation, the only remaining values to calculate in Eq.~\ref{eq:Sij} are $|d\vec\mu/dQ_1|$ and $|d\vec\mu/dQ_2|$, both evaluated at the equilibrium positions for the vibrational mode. In our work, we use the values and uncertainty estimates from theory done with Lan Cheng and Chaoqun Zhang of Johns Hopkins University, and partially presented in Ref.~\cite{Vilas2023_BlackbodyLifetimes}. 

Using these formulas, we estimate the spontaneous and BBR lifetimes for each of the science states, $\X(200), ~\X(03^10),\text{ and }\X(010)$. We add these to the estimated vacuum lifetime to get the resultant "full" lifetime for each state, as given in Table~\ref{tab:lifetimes}. We add them as rates  
\begin{equation*}
    1/\tau_\text{full} = 1/\tau_\text{sp} + 1/\tau_\text{BBR} + 1/\tau_\text{vac},
\end{equation*}
and use a vacuum lifetime of $\tau_\text{vac}\approx3$~s, estimated from the vacuum lifetime in systems in our lab with similar vacuum and consistent with the measured loss rate of molecules in our ODT ($\tau$), see below. The range of this estimated value of $\tau_\text{vac}$ does not significantly affect our results.

{\bf{Comparison}}---The estimated theoretical lifetimes for the $\X(200)$ and $\X(03^10)$ states agree with the measured values within the experimental 1$\sigma$.  The agreement with the $\X(010)$ state lifetime is at the experimental 3$\sigma$ level. This $\sim25$\% lifetime difference could be accounted for by uncertainties in $|d\vec\mu/dQ_1|$ and $|d\vec\mu/dQ_2|$, where %
calculation uncertainties of $10-30$\% are expected. 


\begin{table}[]
    \centering
    \resizebox{\columnwidth}{!}{%
    \begin{tabular}{|c|c|c|c|c|}
    \hline
    \textbf{State} & \textbf{\begin{tabular}[c]{@{}c@{}}Spontaneous decay\\  lifetime (ms)\end{tabular}} & \textbf{\begin{tabular}[c]{@{}c@{}}BBR\\ lifetime (ms)\end{tabular}} & \textbf{\begin{tabular}[c]{@{}c@{}}Full estimated\\  lifetime (ms)\end{tabular}} & \textbf{\begin{tabular}[c]{@{}c@{}}Measured\\  value (ms)\end{tabular}} \\ \hline \hline
    $\X(200)$ & 178 & 699 & 135 & \ODTLifetimeDMStretch(\ODTLifetimeDMStretchunc) \\ \hline
    $\X(010)$ & 1117 & 896 & 427 & \ODTLifetimeEDM(\ODTLifetimeEDMunc) \\ \hline
    $\X(03^10)$ & 408 & 528 & 214 & \ODTLifetimeDMBend(\ODTLifetimeDMBendunc) \\ \hline
    \end{tabular}
    }
    \captionsetup{justification=justified, singlelinecheck=false}
    \caption{\justifying Theory estimated and measured lifetimes for the science states in SrOH. The full estimated lifetime uses the vacuum lifetime of 3~s.}
    \label{tab:lifetimes}
\end{table}

As mentioned above, we measure the trap lifetime molecules in the ODT just after loading, without further state preparation. Due to the low scattering rate of SF cooling most molecules occupy the $\X(000)$ state, we estimate $\gtrsim90$\%, 
so the trap lifetime is greatly dominated by the $\X(000)$ lifetime. We measure this trap lifetime to be $\tau =\ODTLifetimeMain(\ODTLifetimeMainunc)$~s. Since the $\X(000;N=1)$ state is only 15~GHz above the absolute ground state of the molecule, $\X(000;N=0^+)$, the spontaneous lifetime is very large and not considered here. The BBR limited lifetime of $\X(000)$ was considered in Ref~\onlinecite{Vilas2023_BlackbodyLifetimes} and estimated to be 1.3~s, dominated by transitions to three other vibrational states. Because BBR transitions are E1 transitions, they flip parity and, thus, a BBR transition from $\X(000; N=1)$ to another vibrational state would go to a positive parity state, which is not repumped in our optical cycle. Therefore, we expect that $\tau$
is limited by the BBR lifetime from $\X(000; N=1)$ and find our measured value to be within 2$\sigma$ of the estimated $\X(000)$ BBR lifetime \cite{Vilas2023_BlackbodyLifetimes}.







\end{document}